\documentclass[]{ptptex}
\usepackage{graphicx}
\usepackage{bm}

\def\etal{{\it et al.\ }}
\def\beq{\begin{equation}}
\def\eeq{\end{equation}}
\def\pa{\partial}
\def\ra{\rightarrow}
\def\tt{\tilde{t}}
\def\tr{\tilde{r}}
\def\tp{\tilde{\phi}}
\def\tv{\tilde{V}}
\def\tm{\tilde{m}}
\def\to{\tilde{\omega}}
\def\tE{\tilde{E}}
\def\tQ{\tilde{Q}}

\begin{document}

\title{Stability of Q-balls and Catastrophe}
\author{Nobuyuki \textsc{Sakai}
\footnote{nsakai@e.yamagata-u.ac.jp}
and Misao \textsc{Sasaki}\footnote{misao@yukawa.kyoto-u.ac.jp}}
\inst{Department of Education, Yamagata University, Yamagata 990-8560, Japan\\
Yukawa Institute for Theoretical Physics, Kyoto University, Kyoto 990-8502, Japan}
\recdate{April 15, 2007}

\abst{
We propose a practical method for analyzing stability of Q-balls for 
the whole parameter space, which includes the intermediate region between 
the thin-wall limit and thick-wall limit as well as Q-bubbles 
(Q-balls in false vacuum), using catastrophe theory.
We apply our method to the two concrete models, 
$V_3=m^2\phi^2/2-\mu\phi^3+\lambda\phi^4$ and 
$V_4=m^2\phi^2/2-\lambda\phi^4+\phi^6/M^2$.
We find that $V_3$ and $V_4$ Models fall into 
{\it fold catastrophe\/} and {\it cusp catastrophe}, respectively,
and their stability structures are quite different from each other.
}


\maketitle

Q-balls \cite{Col85}, a kind of non-topological solitons \cite{LP92}, 
appear in a large family of  
field theories with global U(1) (or more) symmetry, 
and could play an important role in cosmology. 
For example, the Minimal Supersymmetric Standard Model may contain 
baryonic Q-balls, 
which could be responsible for baryon asymmetry \cite{EM98} and dark 
matter \cite{KS98}.

The stability of Q-balls has been studied in the literature.
Coleman argued that Q-balls are absolutely stable if the charge $Q$ is
sufficiently large, using the thin-wall approximation \cite{Col85}. 
Kusenko showed that Q-balls with small $Q$ are also stable for the 
potential 
\beq\label{V3}
V_3(\phi)={m^2\over2}\phi^2-\mu\phi^3+\lambda\phi^4 ~~~
{\rm with} ~~~ m^2,~\mu,~\lambda>0,
\eeq
using the thick-wall approximation \cite{Kus97a}. Here the thick-wall limit is 
defined by the limit of $\omega^2\rightarrow m^2$, where $\omega$ is the 
angular velocity of phase rotation.
Multamaki and Vilja found that in the thick-wall limit the stability 
depends on the form of the potential \cite{MV00}. Paccetti Correia 
and Schmidt showed a useful theorem which applies to any equilibrium 
Q-balls \cite{PCS01}: 
their stability is determined by the sign of $(\omega/Q)dQ/d\omega$.

It is usually assumed that the potential has an absolute minimum at $\phi=0$.
If $V(0)$ is a local minimum and the absolute minimum is located at $\phi\ne0$,
true vacuum bubbles may appear \cite{Col77}. 
If $Q=0$, vacuum bubbles are unstable: either expanding or contracting.
Kusenko \cite{Kus97b} and Paccetti Correia and Schmidt \cite{PCS01} showed, 
however, that there are stable bubbles if $Q\ne0$. 
They called those solutions ``Q-balls in the false vacuum". 
Hereafter we simply call them ``Q-bubbles".

The standard method for analyzing stability is to take the second variation
of the total energy (given by Eq.(\ref{E}) below) and evaluate its sign.
However, this calculation can be executed analytically only for some 
limited cases; in general the eigenvalue problem should be solved numerically,
as Axenides \etal did \cite{AKPF00}.
In this paper, we propose an easy and practical method for analyzing 
stability with the help of catastrophe theory. The basic idea of
catastrophe theory is described in Appendix.
As we shall show below, once we find {\it behavior variable}({\it s}), 
{\it control parameter}({\it s}) and a {\it potential\/} in the Q-ball system, 
it is easy to understand the stability structure of Q-balls for 
the whole parameter space including the intermediate region 
between the thin-wall limit and thick-wall limit as well as Q-bubbles.

Consider an SO(2)-symmetric scalar field, whose action is given by
\beq\label{S}
{\cal S}=\int d^4x\left[
-\frac12\eta^{\mu\nu}\{\pa_{\mu}\phi_1\pa_{\nu}\phi_1
+\pa_{\mu}\phi_2\pa_{\nu}\phi_2\}
-V(\phi) \right],
~~~{\rm with}~~~
\phi\equiv\sqrt{\phi_1^2+\phi_2^2}.
\eeq
We consider spherically symmetric configurations of the field.
Assuming homogeneous phase rotation,
\beq\label{phase}
(\phi_1,\phi_2)=\phi(r)(\cos\omega t,\sin\omega t),
\eeq
the field equation becomes
\beq\label{fe}
{d^2\phi\over dr^2}=-\frac2r{d\phi\over dr}-\omega^2\phi+{dV\over d\phi}\,.
\eeq
This is equivalent to the field equation for a single static scalar 
field with the potential $V_{\omega}\equiv V-\omega^2\phi^2/2$.
Due to the symmetry there is a conserved charge,
\beq\label{Q}
Q\equiv\int d^3x(\phi_1\pa_t\phi_2-\phi_2\pa_t\phi_1)=\omega I,
~~~{\rm where}~~~
I\equiv\int d^3x~\phi^2.
\eeq

Monotonically decreasing solutions $\phi(r)$ with the boundary conditions,
\beq\label{bc}
{d\phi\over dr}(0)=0,~~~\phi(\infty)=0,
\eeq
exist if min$(V_{\omega})<V(0)$ and $d^2V_{\omega}/d\phi^2(0)>0$,
which is equivalent to
\beq\label{omega}
\omega_{\rm min}^2<\omega^2<m^2
~~~{\rm with}~~~
\omega_{\rm min}^2\equiv{\rm min}\left({2V\over\phi^2}\right),~~~
m^2\equiv{d^2V\over d\phi^2}(0)\,,
\eeq
where we have put $V(0)=0$ without loss of generality.
The two limits $\omega^2\ra\omega_{\rm min}^2$ and $\omega^2\ra m^2$ 
correspond to the thin-wall limit and the thick-wall limit, respectively.
The condition $\omega_{\rm min}^2<m^2$ is not so restrictive because it is 
satisfied if the potential has the form,
\beq\label{V0}
V={m^2\over2}\phi^2-\lambda\phi^n+O(\phi^{n+1})~~
{\rm with} ~~ m^2>0,~\lambda>0,~~n\ge3\,.
\eeq
The total energy of the system for equilibrium solutions is given by
\beq\label{E}
E={Q^2\over2I}+\int d^3x\left\{\frac12\left({d\phi\over dr}\right)^2+V\right\}.
\eeq
Note that the variation of $E$ under fixed $Q$, $\delta E/\delta\phi|_Q=0$, 
reproduces the field equation (\ref{fe}).

Let us discuss how we apply catastrophe theory to
the present Q-ball system. Catastrophe theory is briefly
described in Appendix. 
An essential point is to choose {\it behavior variable}({\it s}),
{\it control parameter}({\it s}) and a {\it potential\/} in the Q-ball 
system appropriately.
For a given potential $V(\phi)$ and charge $Q$, 
 we consider a one-parameter family of perturbed field configurations
$\phi_{\omega}(r)$ near the equilibrium solution $\phi(r)$.
The one-parameter family is chosen to satisfy $I[\phi_{\omega}]=Q/\omega$.
Then the energy is regarded as a function of $\omega$,
$E(\omega)\equiv E[\phi_{\omega}]$. 

Because $dE/d\omega=(\delta E/\delta\phi_{\omega})d\phi_{\omega}/d\omega=0$
 when $\phi_{\omega}$ is an equilibrium solution, $\omega$ 
may be regarded as a {\it behavior variable\/} and 
$E$ as the {\it potential}.
On the other hand, the charge $Q$ and the model parameter(s) of $V(\phi)$
can be given by hand, and therefore should be regarded as {\it control parameters}.
We denote the model parameter(s) by $P_i$ ($i=1,2,\cdots$).
Then we analyze the stability of Q-balls as follows.
\begin{itemize}
\item 
Solve the field equation~(\ref{fe}) with the boundary 
condition~(\ref{bc}) numerically to obtain equilibrium 
solutions $\phi(r)$ for various values of $\omega$ and model parameter(s) $P_i$.
\item 
Calculate $Q$ by (\ref{Q}) for each solution to obtain the 
{\it equilibrium space} $M=\{(\omega,P_i,Q)\}$.
We denote the equation that determines $M$ by $f(\omega,P_i,Q)=0$.
\item 
Find folding points where $\pa P_i/\pa\omega=0$ or $\pa Q/\pa\omega=0$ 
in $M$, which are identical to the stability-change points, 
$\Sigma=\{(\omega,P_i,Q)\,|\,{\pa f/\pa\omega}=0, ~f=0\}$.
\item
Calculate the energy $E$ by (\ref{E}) for equilibrium solutions
around a certain point in $\Sigma$ to find whether the point is 
a local maximum or a local minimum. Then we find the stability 
structure for the whole $M$.
\end{itemize}

Now, using the method devised above, we investigate the stability 
of equilibrium Q-balls.
Because it was shown \cite{PCS01} that in the thick-wall limit Q-balls are stable 
if $n<10/3$ for the potential (\ref{V0}) and unstable otherwise,
we consider two typical models. One is given by (\ref{V3}), 
which we call $V_3$ Model, and the other is given by
\beq\label{V4}
V_4(\phi)={m^2\over2}\phi^2-\lambda\phi^4+{\phi^6\over M^2} ~~~
{\rm with} ~~~m^2,~\lambda,~M^2>0\,,
\eeq
which we call $V_4$ Model.
For $V_3$ Model, rescaling the quantities as
\beq\label{rescale3}
\tt\equiv{\mu\over\sqrt{\lambda}}t,~~ \tr\equiv{\mu\over\sqrt{\lambda}}r,~~
\tp\equiv{\lambda\over\mu}\phi,~~
\tv_3\equiv{\lambda^3\over\mu^4}V_3,~~
\tm\equiv{\sqrt{\lambda}\over\mu}m,~~
\to\equiv{\sqrt{\lambda}\over\mu}\omega,
\eeq
the field equation (\ref{fe}), the potential (\ref{V3}), 
the charge (\ref{Q}) and the energy (\ref{E}) are rewritten as
\beq\label{fe3}
{d^2\tp\over d\tr^2}=-\frac2{\tr}{d\tp\over d\tr}-\to^2\tp+{d\tv_3\over d\tp},~~~
\tv_3={\tm^2\over2}\tp^2-\tp^3+\tp^4,~~~
\tE={\lambda^{\frac32}\over M}E,~~~\tQ=\lambda Q.
\eeq
Similarly, for $V_4$ Model, rescaling the quantities as
\beq\label{rescale4}
\tt\equiv\lambda Mt,~~ \tr\equiv\lambda Mr,~~
\tp\equiv{\phi\over\sqrt{\lambda}M},~~
\tv_4\equiv{V_4\over\lambda^3M^4},~~
\tm\equiv{m\over\lambda M},~~
\to\equiv{\omega\over\lambda M},
\eeq
the field equation (\ref{fe}), the potential (\ref{V4}),
the charge (\ref{Q}) and the energy (\ref{E}) are rewritten as
\beq\label{fe4}
{d^2\tp\over d\tr^2}=-\frac2{\tr}{d\tp\over d\tr}-\to^2\tp+{d\tv_4\over d\tp},~~~
\tv_4={\tm^2\over2}\tp^2-\tp^4+\tp^6,~~~
\tE={E\over M},~~~\tQ={Q\over\lambda}.
\eeq

In both models the system is regarded as a mechanical system with 
the {\it behavior variable} $\to$, the {\it control parameters} $\tm^2$ and $\tQ$,
and the {\it potential} $\tE(\to;\tm^2,\tQ)$.
Because $\to_{\rm min}^2=\tm^2-1/2$, the existing condition~(\ref{omega})
reduces to
\beq
0<\tm^2-\to^2<\frac12\,.
\eeq
The thin-wall and thick-wall limits correspond to 
$\tm^2-\to^2\ra1/2$ and $\tm^2-\to^2\ra0$, respectively.
The condition for ordinary Q-balls, $\to_{\rm min}^2\ge0$,
reduces to $\tm^2\ge1/2$, while that for Q-bubbles,
$\to_{\rm min}^2<0$, to $\tm^2<1/2$.

\begin{figure}
\centerline{\includegraphics[scale=.57]{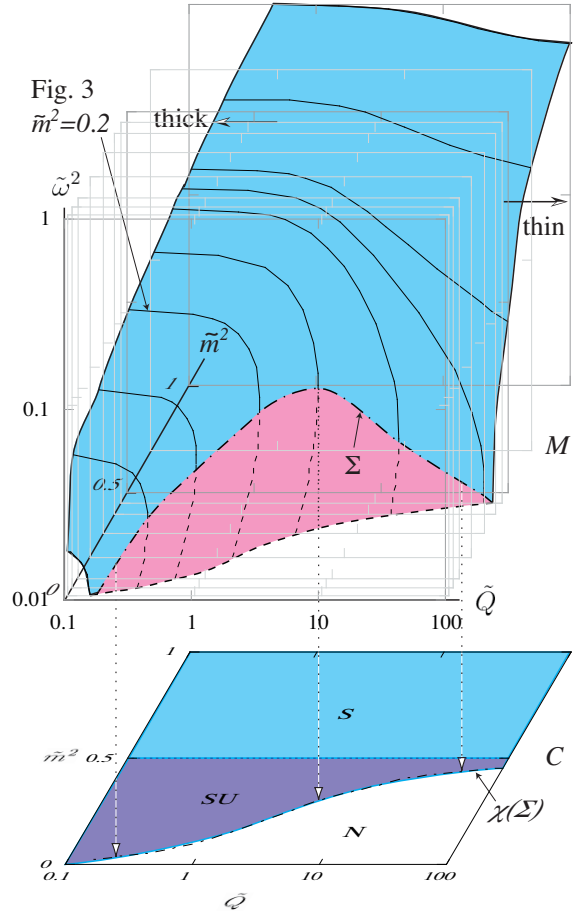}}
\caption{\label{f1}
Structures of the {\it equilibrium spaces},
$M=\{(\to,\tm^2,\tQ)\}$, and their catastrophe map, $\chi(M)$, 
into the {\it control planes}, $C=\{(\tm^2,\tQ)\}$,
for $V_3$ Model.
The dash-dotted lines in $M$ denote stability-change points $\Sigma$,
and the dash-dotted lines in $C$ denote their catastrophe maps $\chi(M)$.
Solid lines in $M$ (on the light-cyan colored surface) and
dashed lines (on the light-magenta colored surface) 
represent stable and unstable solutions, respectively. 
The arrows indicated by ``thin" and ``thick" show
the thin-wall limit, $\to^2\ra\to_{\rm min}^2=\tm^2-1/2$,
and the thick-wall limit, $\to^2\ra \tm^2$, respectively.
In the regions denoted by S, SU and N on $C$,
there are one stable solution, one stable and one unstable solutions,
and no equilibrium solution, respectively, for fixed $(\tm^2,\tQ)$.}
\end{figure}

\begin{figure}
\hspace*{-10mm}\includegraphics[scale=.57]{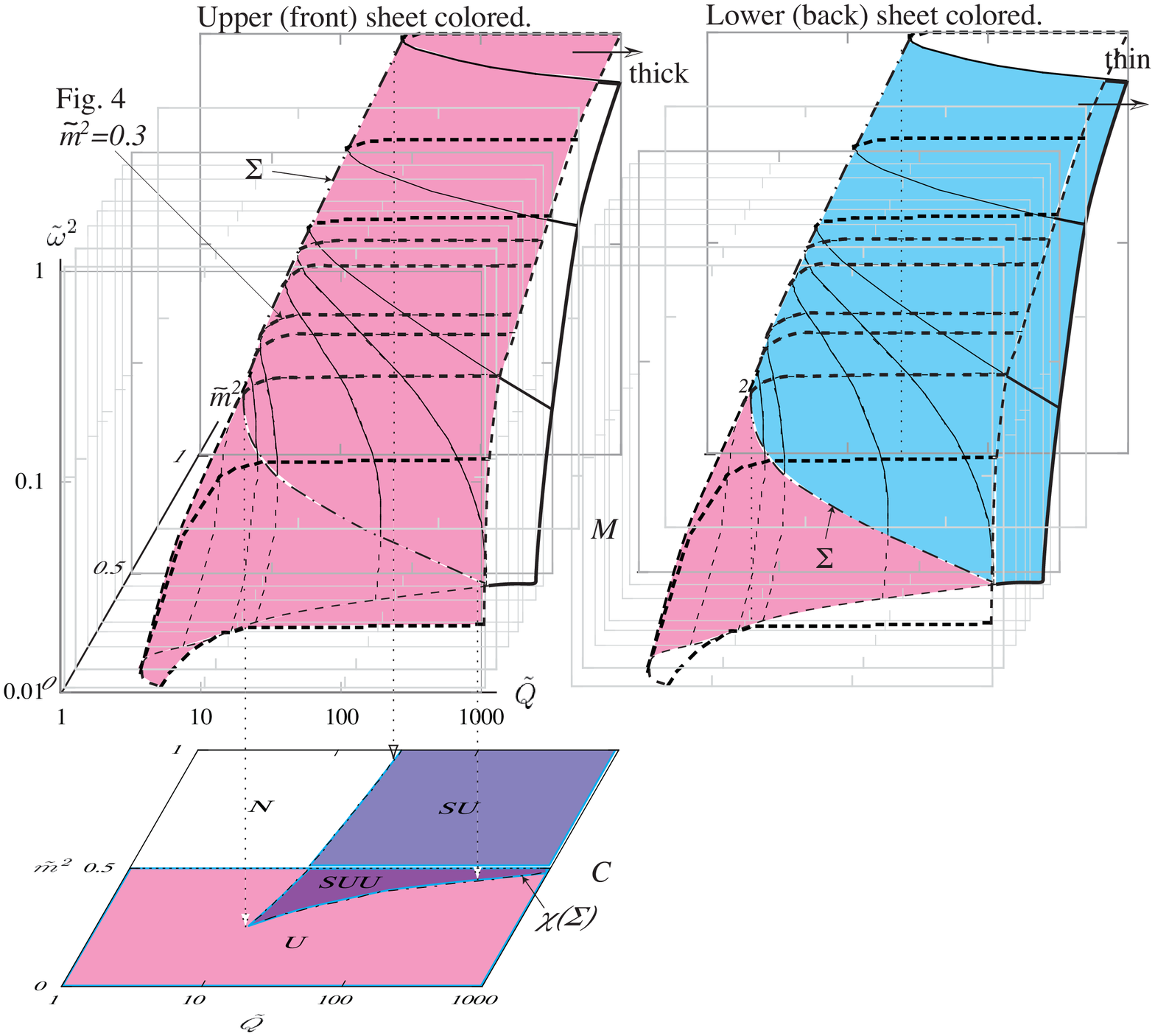}
\caption{\label{f2}
The same as Fig.\ \ref{f1}, but for $V_4$ Model.
Because the structure of $M$ is complicated in this case,
we show two pictures of $M$: The left one shows the
upper (front) sheet of the equilibrium space, while the right one
the lower (back) sheet.
In the regions denoted by N, U, SU and SUU on $C$,
there are no equilibrium solution, one unstable solution, 
one stable and one unstable solutions, and one stable and 
two unstable solutions, respectively, for fixed $(\tm^2,\tQ)$.}
\end{figure}

\begin{figure}
\centerline{\includegraphics[scale=.67]{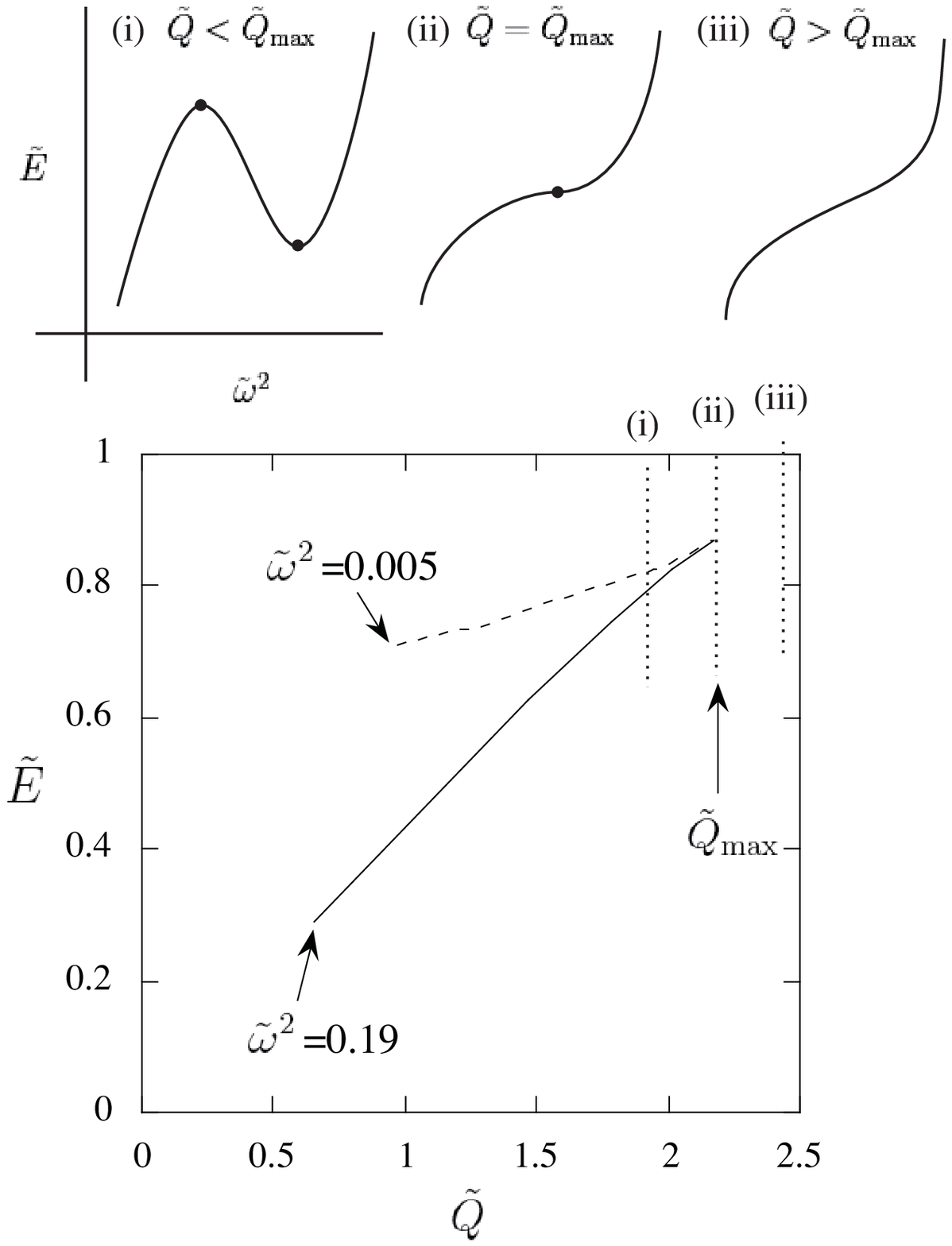}}
\caption{\label{f3}
A schematic picture of the {\it potential} $E(\omega; \tm^2,\tQ)$ 
of $V_3$ Model with $\tm^2=0.2$ near the equilibrium solutions,
and the locus of equilibrium solutions on $(\tQ, \tE)$ plane.
The solid and dashed lines represent stable and unstable
solutions, respectively.}
\end{figure}

\begin{figure}
\includegraphics[scale=.67]{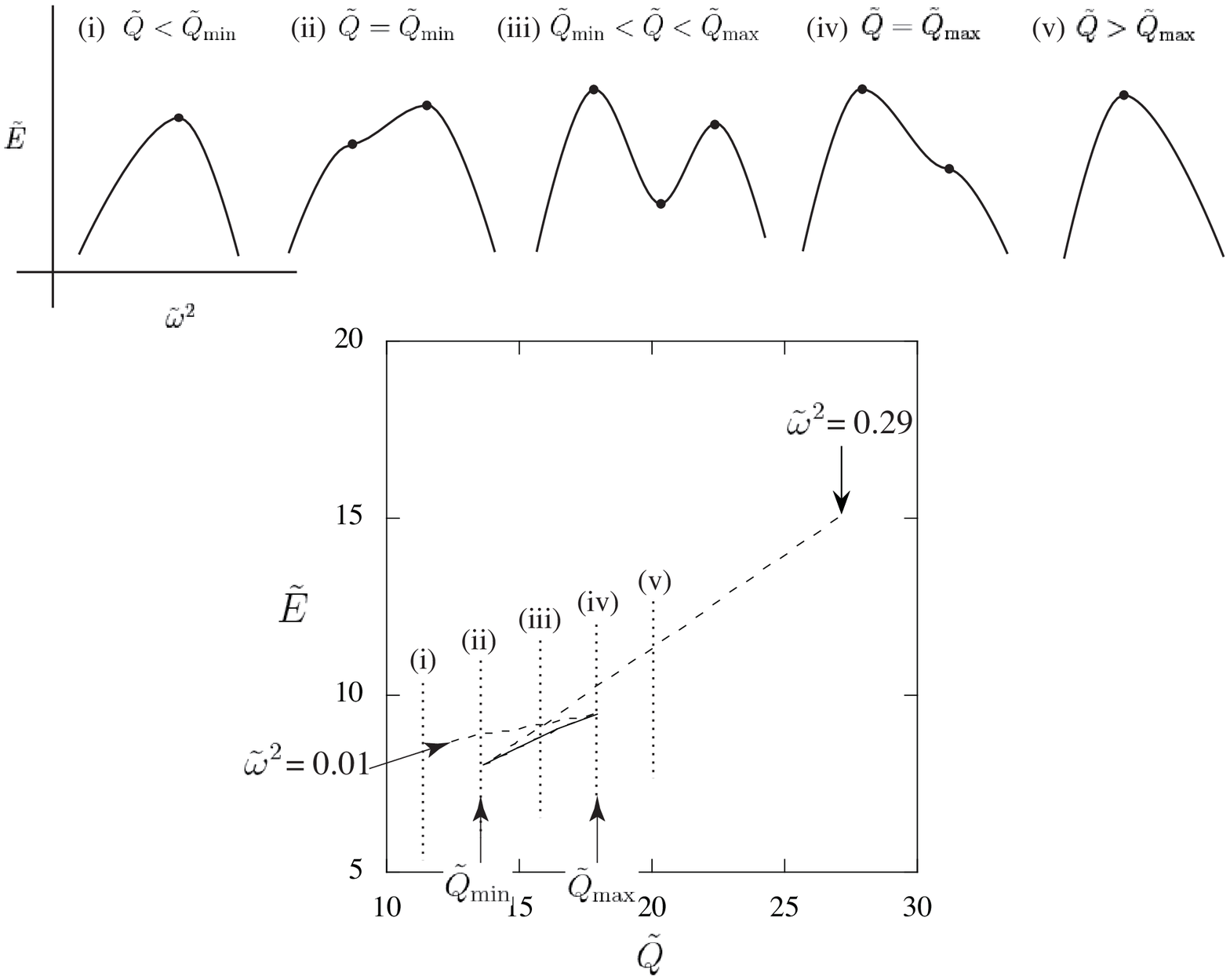}
\caption{\label{f4}
The same as Fig.\ \ref{f3}, but for $V_4$ Model with $\tm^2=0.3$.}
\end{figure}

Figures \ref{f1} and \ref{f2}
show the structures of the {\it equilibrium spaces}, 
$M=\{(\to,\tm^2,\tQ)\}$, and their catastrophe map, $\chi(M)$, 
into the {\it control planes}, $C=\{(\tm^2,\tQ)\}$,
for $V_3$ and $V_4$ Models, respectively. 
We only show the results for $\to>0$; the sign transformation 
$\to\ra-\to$ changes nothing but $\tQ\ra-\tQ$.
The dash-dotted lines in $M$ denote stability-change points $\Sigma$.
Because the equilibrium space alone does not tell us which lines,
solid or dashed, represent stable solutions, 
we evaluate the energy $\tE$ for several equilibrium solutions,
as shown in Figs.\ \ref{f3} and \ref{f4}.
When there are double or triple values of $\tE$ for a given set 
of the control parameters $(\tm^2,\tQ)$, by energetics the solution 
with the lowest value of $\tE$ should be stable and the others should be unstable.
In Figs.~\ref{f3} and \ref{f4}, we also give a sketch of the {\it potential} 
$E(\omega; \tm^2,\tQ)$ near the equilibrium solutions.
Once the stability for a given set of the parameters $(\tm^2,\tQ)$ is found,
the stability for all the sets of parameters which may be reached continuously
from that set without crossing $\Sigma$ is the same. 
We therefore conclude that, in Figs.\ \ref{f1} and \ref{f2}
as well as in Figs.\ \ref{f3} and \ref{f4},
solid and dashed lines correspond to stable and unstable solutions, 
respectively.

According to the configurations of $\chi(\Sigma)$ in the 
{\it control planes\/} in Figs.\ \ref{f1} and \ref{f2},
we find that $V_3$ Model falls into {\it fold catastrophe}
while $V_4$ Model falls into {\it cusp catastrophe}.
In the {\it control planes}, the numbers of stable and unstable solutions 
for each $(\tm^2,\tQ)$ are represented by N, S, U, SU and SUU
 (see the figure captions for their definitions).
Thus we find the stability structures of the two models
 are very different from each other. They are found as follows.

\begin{list}{}{}
\item[$\bm V_3$ \bf Model]~
\begin{list}{$\bullet$}{}
\item $\tm^2\ge1/2$: All equilibrium solutions are stable. 
\item $\tm^2<1/2$ (Q-bubbles): For each $\tm^2$ there is a maximum charge, 
$\tQ_{\rm max}$, above which equilibrium solutions do not exist. 
For $\tQ<\tQ_{\rm max}$, stable and unstable solutions coexists. 
It is interesting to note that stable Q-bubbles exist no matter how small $\tQ$ is.
\end{list}

\item[$\bm V_4$ \bf Model]~
\begin{list}{$\bullet$}{}
\item $\tm^2\ge1/2$: For each $\tm^2$ there is a 
minimum charge, $\tQ_{\rm min}$, below which equilibrium solutions do not exist.
For $\tQ>\tQ_{\rm min}$, stable and unstable solutions coexists.
\item $\tm^2<1/2$ (Q-bubbles): For each $\tm^2$ there is a maximum charge, 
$\tQ_{\rm max}$, as well as a minimum charge, $\tQ_{\rm min}$, 
where stable solutions do not exist if $\tQ<\tQ_{\rm min}$ 
or $\tQ>\tQ_{\rm max}$.
For $\tQ_{\rm min}<\tQ<\tQ_{\rm max}$, there are one stable
and two unstable solutions. 
\\
As $\tm^2$ becomes smaller, $\tQ_{\rm max}$ and $\tQ_{\rm min}$ come 
close to each other, and finally merge at $\tm^2\approx0.26$, 
below which there is no stable solution.
\end{list}
\end{list}
The above results for the two models are consistent with the previous 
results for some special cases such as the thin-wall limit, the thick-wall 
limit and bubbles with $Q=0$.

Although we have investigated only two concrete models, 
taking account of the fact that the stability structure falls into 
two classes in the thick-wall limit, that is, the fact that Q-balls are 
stable if $n<10/3$ for the potential (\ref{V0}) and unstable otherwise,
one expects that there are essentially two distinct stability
structures in the general case. Then the two types of models investigated
here, $V_3$ and $V_4$, may be regarded as the representatives of
these two distinct stability structures.

For example, in the gravity-mediated supersymmetry breaking model \cite{SUSY},
the lowest-order negative term of the potential is $\sim-\phi^2\log\phi$.
Because this term corresponds to $n<3$ in (\ref{V0}), the stability 
structure of this model falls into $V_3$ Type.
Furthermore, because the potential is positive everywhere, which 
corresponds to $\tm^2>1/2$ in Fig.\ 1, all equilibrium solutions are 
stable in this model.

In summary, we have proposed a new method for analyzing 
the stability of Q-balls using catastrophe theory.
An essential point is that, although the Q-ball system (\ref{S}) includes 
infinite degrees of freedom, practically it can be regarded as a mechanical
system with one variable, $\omega$, near equilibrium solutions.
Therefore, we have applied catastrophe theory, which was
 established for mechanical systems with finite degrees of freedom, 
to the Q-ball system.
A similar analysis but on the stability of exotic black holes
was done by Maeda \etal \cite{Maeda} some time ago, and catastrophe theory
was found to be very useful. Thus it seems worthwhile to consider
the application of catastrophe theory to other
 cosmological (gravitating) solitons such as gravitating Q-balls \cite{GQ},
 topological defects, and branes.
It may be also interesting to apply the catastrophe-theoretic approach to
non-relativistic atomic Bose-Einstein condensates \cite{BEC}, where 
Q-ball-like solitons appear. 

We thank H. Kodama, K. Maeda, K. Nakao, V. Rubakov, H. Shinkai,
T. Tanaka and S. Yoshida for useful discussions.
A part of this work was done while NS was visiting at Yukawa 
Institute for Theoretical Physics, which was supported by Center for 
Diversity and Universality in Physics (21COE) in Kyoto University.
The numerical computations of this work were carried out at the Yukawa 
Institute Computer Facility.
This work was supported in part by JSPS Grant-in-Aid for Scientific Research
(B) No.\ 17340075, (A) No.\ 18204024 and (C) No.\ 18540248.

\appendix
\section{Basic Idea of Catastrophe Theory}

To illustrate the basic idea of catastrophe theory \cite{PS78}, 
we consider a system with one {\it behavior variable} $x$, two 
{\it control parameters} $p,q$ and  a {\it potential} $F(x;p,q)$.
An equilibrium point of $x$ is determined by $dF/dx=0$ for 
each pair of $(p,q)$. 
The set of the control parameters, $C\equiv{(p,q)}$, spans a plane
called the {\it control plane}, and the set of equilibrium points,
\beq
M\equiv\left\{(x,p,q)|f(x,p,q)\equiv {dF\over dx}=0\right\},
\eeq
is called the {\it equilibrium space}.
Because equilibrium points are stable if $\pa f/\pa x>0$, the 
boundary of stable and unstable equilibrium points are given by the curve,
\beq
\Sigma\equiv\left\{(x,p,q)|{\pa f\over\pa x}=0, ~f=0\right\}.
\eeq

The {\it catastrophe map} is defined as
\beq
\chi:~M\ra C,~~(x,p,q)\ra(p,q).
\eeq
According to Thom's theorem, depending on the configurations of the image $\chi(\Sigma)$,
all mechanical systems with stability-change are classified into several catastrophe types.
If the number of control parameters is two, as is this example, possible catastrophe types are {\it fold catastrophe\/} and  {\it cusp catastrophe}.
As we show in the text, Q-ball models are also classified into these two types.

If the {\it potential} $F(x;p,q)$ is known, it is easy to find equilibrium 
points and their stability. However, even if we do not know
the explicit form of $F(x;p,q)$, 
we can still find $\Sigma$ by analyzing equilibrium points as follows. 
The Taylor expansion of $f(x,p,q)$ in the vicinity of a certain point
 P$(x_0,p_0,q_0)$ in $M$, where $f=0$, up to the first order yields
\beq\label{qex}
q=q(x,p)=q_0-\left({\pa f\over\pa q}\right)^{-1}\left\{{\pa f\over\pa x}(x-x_0)
+{\pa f\over\pa p}(p-p_0)\right\},
~~~{\rm if} ~~~ {\pa f\over\pa q}\ne0.
\eeq
Because $\pa f/\pa x=0$ in $\Sigma$, it follows from (\ref{qex}) 
that $\pa q/\pa x=0$ in $\Sigma$. 
Similarly, unless $\pa f/\pa x=0$, $\pa p/\pa x=0$ in $\Sigma$.
Therefore, surveying the points with $\pa p/\pa x=0$ or $\pa q/\pa x=0$ 
in the {\it equilibrium space} $M$, we can obtain the set of stability-change
 points $\Sigma$.

\end{document}